\documentclass[12pt,draftclsnofoot, onecolumn]{IEEEtran}
\usepackage{bbm}
\usepackage{amsfonts}
\usepackage{multirow}
\usepackage{mathrsfs}
\usepackage{graphicx,cite,epsfig,amssymb,amsmath}

\begin{document}

\title{Single-RF MIMO: From Spatial Modulation to Metasurface-Based Modulation}


\author{\normalsize
Qiang Li$^{\star}$, Miaowen Wen$^\star$, and Marco Di Renzo$^\#$\\
$^\star$School of Electronic and Information Engineering\\
South China University of Technology, Guangzhou 510640, China\\
Email: eeqiangli@mail.scut.edu.cn; eemwwen@scut.edu.cn\\
$^\#$Laboratoire des Signaux et Syst\`{e}mes, CNRS, CentraleSup\'{e}lec,\\
Universit\'{e} Paris-Saclay, 91192 Gif-sur-Yvette, France\\
Email: marco.direnzo@centralesupelec.fr}
\vspace{+0.8cm}

\maketitle

\begin{abstract}
In multiple-input multiple-output (MIMO), multiple radio frequency (RF) chains are usually required to simultaneously transmit multiple data streams. As a special MIMO technology, spatial modulation (SM) activates one transmit antenna with one RF chain and exploits the index of the active antenna for information transfer at each time slot. Recently, reconfigurable metasurfaces have emerged as a promising technology that is able to reconfigure the wireless propagation environment by altering the amplitude and/or phase of the incident signal. In this article, we aim for the implementation of single-RF MIMO by shifting the focus from SM to metasurface-based modulation. Specifically, the principles of SM and metasurfaces are first presented. After reviewing the evolution of SM, we elaborate on the idea of metasurface-aided single-RF MIMO and discuss some implementations for it in line with notable variants of SM. A comparison between antenna-based and metasurface-based modulation is made to highlight the advantages of using metasurfaces. We finally investigate the research challenges and opportunities in the context of metasurface-based modulation.
\end{abstract}


\newpage

\IEEEpeerreviewmaketitle

\section*{Introduction}
Multiple-input multiple-output (MIMO) is a ubiquitous wireless technology that is able to multiply the link capacity by using multiple transmit and receive antennas. In general, however, a number of radio frequency (RF) chains equal to the number of transmit antennas are required to simultaneously deliver multiple data streams. The requirement of multiple RF chains complicates the system implementation, expands the hardware cost, and increases the power consumption. This conventional MIMO technology may not be a good choice for energy-constrained and size-limited devices that intend to increase the data rate.

As the progenitor and a prominent member of index modulation \cite{Access:IM_Survey,Springer:IM}, spatial modulation (SM) appears as a promising and competitive alternative to conventional MIMO. Based on the MIMO framework, SM activates a single transmit antenna for each transmission of a constellation symbol \cite{PIEEE:SM_Survey,ST:SM}. The index of the active antenna is also exploited to convey information thanks to the uniqueness and randomness properties of the wireless channel. Since only a single RF chain is required at the transmitter and the active antenna index carries additional information, SM achieves an appealing trade-off between the spectral efficiency (SE) and energy efficiency in comparison with conventional MIMO and single-antenna systems. Resorting to the SM philosophy, several researchers have developed numerous variants of the SM technology, which typically include generalized SM (GSM), quadrature SM (QSM), receive SM (RSM), and receive QSM (RQSM) \cite{JSAC:SM}. However, SM and its variants still face some challenges.
\begin{itemize}
\item Not all the SM variants inherit the single-RF property of SM. Multiple RF chains are still necessary for RSM and RQSM. Furthermore, the positions of the active transmit antennas dynamically change for each channel use in SM, GSM, and QSM. As a consequence, a high-speed RF switch that operates at the symbol rate and introduce low switching losses plays a vital role in the implementation of single-RF SM/GSM/QSM transmitters. These issues increase the system complexity and implementation cost.

\item The SE of SM members benefits from the increase in the number of antennas. However, it is not a straightforward task to design a compact transceiver that can support a large number of sufficiently distinguishable antennas. On the other hand, due to the limited physical size of transceivers, there may be high correlation among different antennas for SM members. This obscures the distinctness of antenna states, and thus results in performance degradation.

\item SM-based solutions are based on the fact that different active antennas lead to different channel realizations, whereas they cannot configure the wireless propagation environment itself. The system performance of SM highly depends on the distinctness of the channel signatures associated with different active antennas. Hence, rich scattering in the propagation environment and/or a large number of receive antennas are required for SM to avoid poor error performance. Unfortunately, these requirements may not be satisfied in general.
\end{itemize}

It is worth mentioning that only the activation state of an \textit{ordinary} antenna is utilized for spatial information transfer in SM and its variants. \textit{Reconfigurable} antennas, on the other hand, increase the degrees of freedom for modulating and encoding data. Besides the activation state, in particular, the radiation pattern, the polarization state, and the frequency of operation, that can be altered intentionally and can be used for transmitting information \cite{ACCESS:SM_RA}. This observation has motivated researchers to consider the idea of using reconfigurable antennas for index modulation. Early attempts include SM with reconfigurable antennas \cite{ACCESS:SM_RA} and media-based modulation (MBM) \cite{WCM:MBM}, both of which embed information into the radiation patterns of reconfigurable antennas, and enable a single-RF transmitter with one transmit antenna. From a broad point of view, MBM can be viewed as a special instance of metasurface-based modulation. The technology of metasurfaces, a.k.a. reconfigurable intelligent surfaces (RISs), has recently emerged as a game-changer for future wireless communication networks \cite{JSAC:RIS_T}. An RIS is a man-made planar surface of electromagnetic materials whose characteristics can be configured by an external controller. The most distinctive property of RISs lies in making the environment controllable and empowering the concept of smart radio environments \cite{JWCN:RIS_Survey}. With the aid of RISs, an uncontrollable and unfavorable environment can be turned into a controllable and benign entity.

In wireless networks, an RIS is often designed to behave as a programmable electromagnetic reflector with or without transmitting its own information \cite{WC:RIS_MIMO}. Different from relay-aided communications, an RIS has the capability of overcoming the negative effects of electromagnetic propagation through reconfiguring incident signals without using a power amplifier, increasing the background noise, and generating new signals \cite{RIS_Relay}. Therefore, RISs provide numerous opportunities for single-RF transmissions. The challenges faced by SM and the advantages of RISs motivate this article to shift the focus of single-RF MIMO design from SM to metasurface-based modulation. In this article, first, we present the principles of SM and RISs, and then review the variants of SM; afterward we introduce the implementation of single-RF MIMO from the perspective of metasurface-based modulation, followed by a comparison between antenna-based and metasurface-based modulation; finally, we point out possible research directions and potential challenges on metasurface-based modulation.

\section*{Principles of SM and Metasurfaces}
In this section, the principles of SM and metasurfaces are presented as preliminaries.
\subsection*{Principle of SM}
\begin{figure}[t]
	\centering
	\includegraphics[width=4.5in]{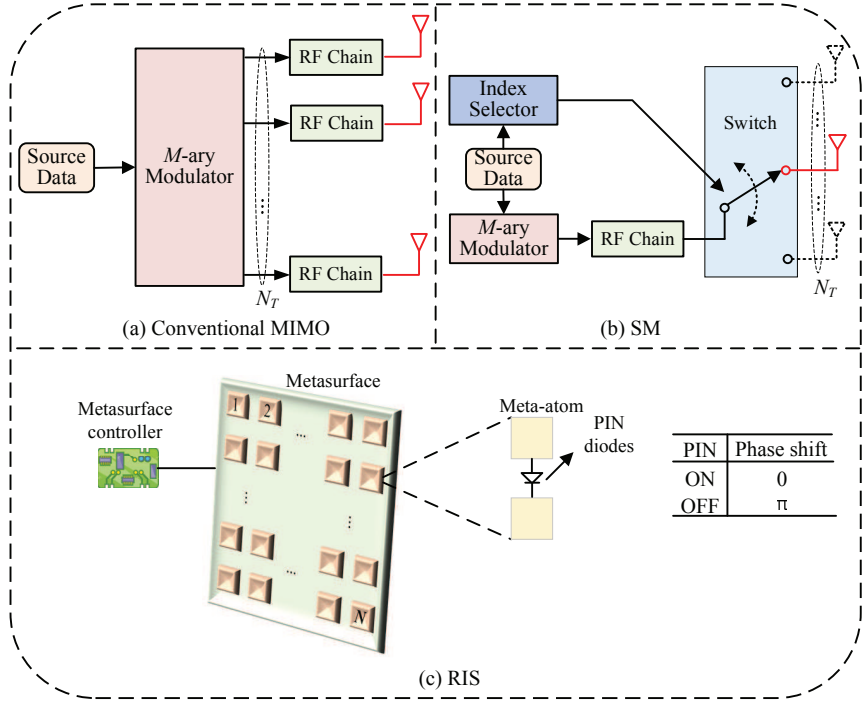}
	\caption{Architectures of (a) conventional MIMO, (b) SM, and (c) an RIS.}
	\label{Fig1}
\end{figure}
In Figs.~\ref{Fig1}(a) and (b), we make a comparison between the transmitter structures of conventional MIMO and SM, both equipped with $N_T$ transmit antennas. For conventional MIMO, all $N_T$ transmit antennas are activated to deliver $N_T$ data streams simultaneously. Therefore, $N_T$ RF chains are needed. In SM, by contrast, there is only a single RF chain and a single active antenna for each channel use. The index of the active antenna is determined by the information to be transmitted, and the RF chain is dynamically switched to that antenna. Specifically, the source data is divided into two parts. The first part, termed spatial information, is loaded into the index selector to determine the active antenna index. Since there are $N_T$ possible realizations of the active antenna index, $\lfloor \log_2 N_T\rfloor$ bits can be conveyed by the active antenna selection, where $\lfloor \cdot \rfloor$ is the floor function. As an example, for $N_T = 2$, one bit is conveyed by the active antenna index. We can simply take the rule that ``0'' and ``1'' imply that the first and second antennas are activated, respectively. The second part with $\log_2 M$ bits is mapped to an $M$-ary symbol that is further transmitted from the selected (active) antenna. The SM receiver not only detects the $M$-ary symbol, but also ascertains which transmit antenna is active. SM enjoys low hardware complexity and power consumption since a single RF chain is employed.

\subsection*{Principle of Metasurfaces}
An RIS consists of a large number of low-cost and passive meta-atoms (elements). Each element is coupled with electronic devices such as positive-intrinsic-negative (PIN) diodes whose signal response (reflection amplitude and phase shift) can be adjusted in real time by the stimuli from the external world. Fig.~\ref{Fig1}(c) depicts a typical architecture of an RIS, where $N$ elements are printed on a dielectric substrate, and a smart controller is attached to the RIS. The metasurface controller tunes the reflection amplitude/phase shift of each element and coordinates with other network components through separate low-rate wire/wireless links. An example of a meta-atom is also shown in Fig.~\ref{Fig1}(c), where an embedded PIN diode can impose $0$ or $\pi$ of phase shift on the incident signal depending on the bias voltage. As such, by coupling more PIN diodes with each meta-atom and carefully designing the connectivity between them, more possible phase shifts can be obtained by setting the corresponding bias voltages. On the other hand, the reflection amplitude of each element can be modified via variable resistors.

By assuming that the meta-atoms are designed and optimized independently of the others (local design), the reflection coefficient for the $n$-th element can be modeled as $\beta_n = a_n e^{j\phi_n}$, where $a_n\in[0,1]$ is the reflection amplitude and $\phi_n \in[0,2\pi)$ denotes the phase shift for $n=1,\ldots,N$. In practice, $a_n$ and $\phi_n$ are discrete values taken from their respective ranges. When a radio wave radiates out towards an RIS, each element of the RIS reflects the incident signal with its reflection coefficient. The equivalent channel from the transmitter to the receiver through the RIS is the composition of $N$ cascaded channels, each of which includes the incident link, the element's reflection, and the reflecting link. Hence, by adjusting the reflection coefficients to cater to dynamic wireless channels, RISs can control the reflected signal to achieve different purposes. For instance, to enhance the received power of the intended user, we can set $a_n=1$ for all $n$ and take the values of $\phi_n$ that achieve phase alignment; whereas to convey the RIS'private information, we can change the values of $a_n$ and/or $\phi_n$ according to the information bits to be transmitted.

\section*{Evolution of SM}
\begin{figure}[t]
	\centering
	\includegraphics[width=4.0in]{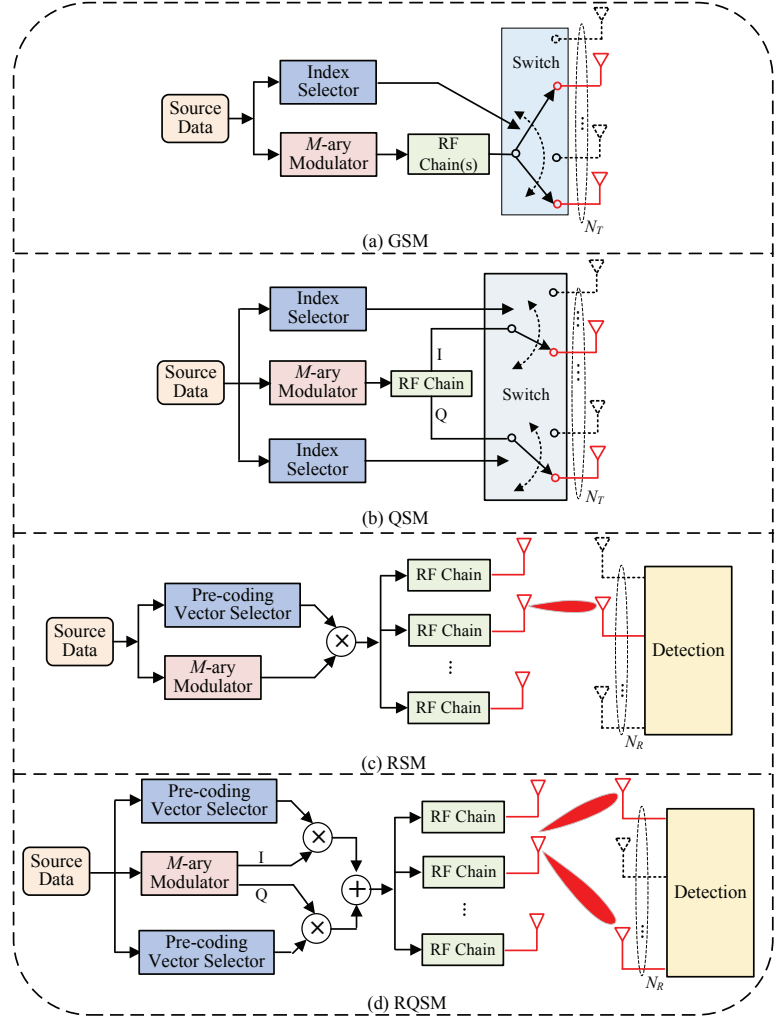}
	\caption{The evolution of SM: (a) GSM, (b) QSM, (c) RSM, and (d) RQSM.}
	\label{Fig2}
\end{figure}
Since the SE of SM suffers from a slow logarithmic growth with an increasing number of transmit antennas, enhanced variants of SM with higher SE have been proposed. Moreover, the SM philosophy has been applied to receive antennas. In this section, the evolution of SM is outlined by discussing four representative variants of SM, namely GSM, QSM, RSM, and RQSM, whose transmitter structures are shown in Fig.~\ref{Fig2}. In order to highlight the differences among the implementations of antenna-based and metasurface-based single-RF MIMO, we focus only on the transmitter design unless otherwise specified.

\subsection*{GSM}
The transmitter structure of GSM is shown in Fig.~\ref{Fig2}(a), where $N_A$ transmit antennas, rather than one as in SM, are activated from $N_T$ available antennas according to $\lfloor \log_2{N_T \choose N_A} \rfloor$ information bits for each channel use, where $N_A \in \{2,\ldots, N_T-1\}$ and ${\cdot \choose \cdot}$ denotes the binomial coefficient. Depending on whether the $N_A$ active antennas convey the same or different constellation symbols, GSM can be implemented with one or multiple RF chains. For single-RF GSM, one $M$-ary symbol is loaded into the RF chain and the resulting RF signal is transmitted from the $N_A$ active antennas, achieving an SE of $\lfloor \log_2{N_T \choose N_A} \rfloor + \log_2 M$ bpcu; for multi-RF GSM, $N_A$ different $M$-ary symbols are separately empowered by $N_A$ RF chains and transmitted from the $N_A$ active antennas, achieving an SE of $\lfloor \log_2{N_T \choose N_A} \rfloor + N_A \log_2 M$ bpcu.
\subsection*{QSM}
Fig.~\ref{Fig2}(b) presents the transmitter structure of QSM, which performs SM independently on the in-phase (I-) and quadrature (Q-) dimensions. In QSM, the source data is partitioned into three parts to be transmitted. The first part made of $\log_2 M$ bits is mapped to an $M$-ary symbol, which is further fed into an RF chain. The second and third parts both having $\lfloor\log_2 N_T\rfloor$ bits are used to select the antennas for transmitting the I- and Q- components of the RF signal, respectively. QSM doubles the number of spatial information bits and maintains the single-RF property of SM in spite of necessitating one or two active antennas at each time slot.
\subsection*{RSM}
RSM, a.k.a. pre-coding aided SM, applies the concept of SM at the receiver side. With the aid of transmit pre-coding (e.g., zero-forcing and minimum mean-squared error processing), RSM utilizes the indices of the receive antennas to convey spatial information in addition to the conventional constellation information of $M$-ary symbols. Hence, the SE of RSM is $\lfloor\log_2 N_R\rfloor + \log_2 M$ bits per channel use (bpcu), where $N_R$ is the number of receive antennas. Fig.~\ref{Fig2}(c) shows the system model of RSM, where a constellation symbol is pre-coded by a vector selected from a predefined set of $N_R$ entities. In particular, different pre-coding vectors lead to different targeted receive antennas. In contrast to SM that is characterized by an easy-to-implement transmitter, RSM requires $N_T$ RF chains and channel state information at the transmitter (CSIT). However, RSM enjoys high beamforming gain and low complexity receiver design, which are highly desired for downlink MIMO transmissions.

\subsection*{RQSM}
RQSM, as its name implies, is the amalgamation of QSM and RSM. Fig.~\ref{Fig2}(d) presents the system model of RQSM. Specifically, the I- and Q- parts of an $M$-ary symbol undergo independent beamforming, each of which points to a receive antenna. In RQSM, both pre-coding vectors are selected from a common predefined set according to the source data. For each channel use, RQSM employs the index of a receive antenna twice to convey spatial information. Compared with RSM, RQSM achieves an increase of $\lfloor\log_2 N_R\rfloor$ bpcu in terms of SE.

\begin{figure}[t]
	\centering
	\includegraphics[width=4.6in]{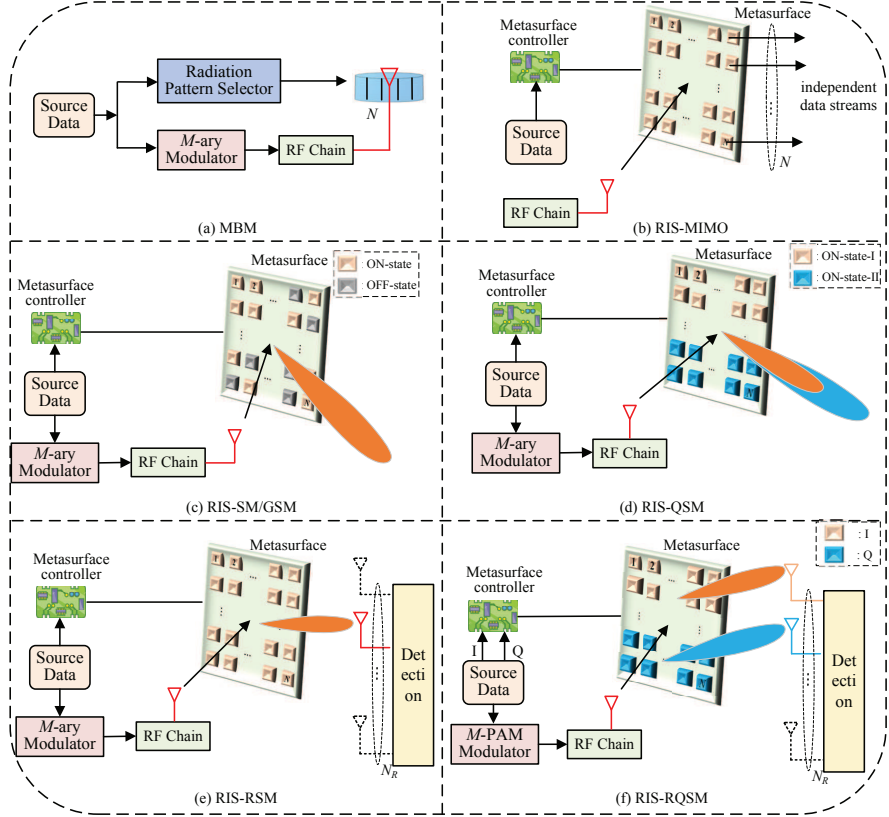}
	\caption{Single-RF metasurface-based modulation schemes: (a) MBM, (b) RIS-MIMO, (c) RIS-SM/GSM, (d) RIS-QSM, (e) RIS-RSM, and (f) RIS-RQSM.}
	\label{Fig3}
\end{figure}
\section*{Metasurface-Based Modulation}
As discussed in previous text, RISs can remodulate an unmodulated/modulated incident signal by adjusting the reflection coefficients of the meta-atoms. This feature greatly facilitates the design of single-RF modulation schemes. In this section, we elaborate on the idea of single-RF metasurface-based modulation via six possible options, including MBM, RIS-MIMO, RIS-SM/GSM, RIS-QSM, RIS-RSM, and RIS-RQSM. The corresponding system models are shown in Fig.~\ref{Fig3}. It is worth noting that the concept of metasurface-based modulation was first introduced in \cite{JWCN:RIS_Survey} via an example that a sensor coated with an RIS transmits its own 1-bit information by differentiating or not altering the reflected signal.

\subsection*{MBM}
Unlike SM that exploits the activation state of regular antennas for spatial information transmissions, MBM utilizes the radiation patterns of reconfigurable antennas to convey spatial information \cite{WCM:MBM}. In MBM, reconfigurable RF mirrors, which act as ON/OFF switches, are deployed around a radiating element. By configuring the status of the switches, different radiation patterns can be created and then employed for index modulation. Fig.~\ref{Fig3}(a) depicts the transmitter structure of MBM with $N$ RF mirrors. Specifically, for each channel use, a part of source data, consisting of $\log_2 M$ bits, is mapped to an $M$-ary symbol and then emitted from the RF chain; the rest of the source data, consisting of $N$ bits, is used to control the ON/OFF status of each RF mirror. Notably, the $N$ RF mirrors can be viewed as a sort of metasurface with $N$ reflecting elements, where the reflection amplitude is $a_n\in \{0,1\}$ and the reflection phase is $\phi_n=0$ for $n=1,\ldots,N$.

\subsection*{RIS-MIMO}
Each element of an RIS can reflect a radio wave impinging upon it by changing its amplitude and phase. Therefore, amplitude and phase modulation, such as phase shift keying (PSK) and quadrature amplitude modulation (QAM), can be easily implemented at each reflecting element. Hence, a single-RF MIMO transmitter can be built by illuminating an RIS with an unmodulated carrier signal and by performing amplitude and phase modulation at each reflecting element according to the source data \cite{WC:RIS_MIMO}, as shown in Fig.~\ref{Fig3}(b). For RIS-MIMO, the number of information streams that can be simultaneously transmitted (via reflections) is equal to the number of meta-atoms of the RIS.

\subsection*{RIS-SM/GSM}
Since the ON (active) and OFF (inactive) status of each reflecting element can be realized by setting the reflection amplitude to be 1 and 0, respectively, the SM/GSM principle can also be applied to an RIS, forming the scheme of RIS-SM/GSM \cite{JSAC:Reflecting,JSAC:RIS_SM,TWC:GSM}. Fig.~\ref{Fig3}(c) gives the structure of an RIS-SM/GSM transmitter. One part of the source data is mapped to an $M$-ary symbol, while the other part is loaded into the metasurface controller to determine the positions of active reflecting elements. Hence, the SE of RIS-SM/GSM is $\log_2 M + \lfloor \log_2 {{N}\choose{N_A}} \rfloor$ bpcu, where $N_A\in\{1,\ldots,N-1\}$ is the number of active elements. Provided that the CSI is available at the RIS, passive beamforming can be carried out to minimize the bit error rate (BER) by tuning the phase shift of each reflecting element. At high signal-to-noise ratio (SNR), a simpler beamformer is to set the phase shifts to be those that achieve phase alignment. However, RIS-SM/GSM underutilizes the RIS since some reflecting elements are kept inactive.

\begin{figure}[t]
	\centering
	\includegraphics[width=4.6in]{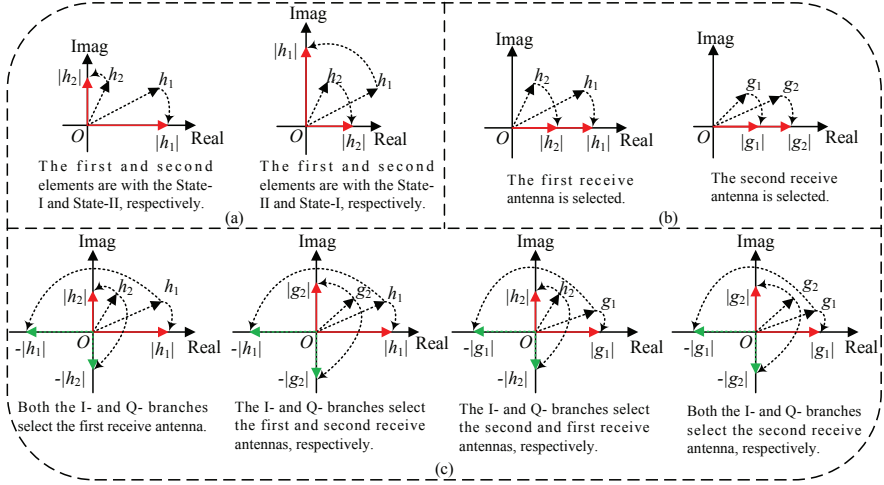}
	\caption{Examples of channel phase shifts for $N=2$: (a) RIS-QSM, where $N_R=1$ and $h_i$ denotes the channel from the $i$-th reflecting element to the receive antenna; (b) RIS-RSM, where $N_R=2$, and $h_i$ ($g_i$) denotes the channel from the $i$-th reflecting element to the first (second) receive antenna; (c) RIS-RQSM, where the first (second) element is controlled by the I- (Q-) branch, $N_R=2$, and $h_i$ ($g_i$) denotes the channel from the $i$-th reflecting element to the first (second) receive antenna.}
	\label{Fig4}
\end{figure}

\subsection*{RIS-QSM}
In contrast to RIS-SM/GSM, RIS-QSM further unlocks the potential of RISs by reactivating the inactive elements, as shown in Fig.~\ref{Fig3}(d). Specifically, in RIS-QSM, the ``active'' and ``re-active'' elements are labeled with the State-I and State-II, respectively, and the positions of ``active'' (``re-active'') elements convey spatial information. The phase shifts for the State-I and State-II are tuned to $-\alpha$ and $-\alpha+\pi/2$, where $\alpha$ is the phase of the channel associated with the receive antenna, as shown for $N=2$ in Fig.~\ref{Fig4}(a). In RIS-QSM, all reflecting elements are employed for beamforming, resulting in improved receive SNR, whereas the SE of RIS-QSM is the same as that of RIS-SM/GSM.

\subsection*{RIS-RSM}
Similar to RSM, RIS-RSM uses the index of a receive antenna as well as an $M$-ary symbol to convey information \cite{TCOM:RIS_RSM}. The SE of RIS-RSM is the same as that of RSM. However, RIS-RSM only needs a single RF chain at the transmitter, and adopts metasurface-based passive beamforming instead of active pre-coding to steer the signal towards a certain receive antenna. Fig.~\ref{Fig3}(e) presents the transmitter design of RIS-RSM, where a modulated carrier signal carrying an $M$-ary symbol radiates out towards an RIS and the reflection coefficients are tuned according to the information bits for maximizing the received power of the selected antenna. Please refer to Fig.~\ref{Fig4}(b) for an example of channel phase shifts with $N=2$ and $N_R=2$.

\subsection*{RIS-RQSM}
By capitalizing on the idea of QSM, RIS-RSM can be extended to RIS-RQSM that uses two independent data streams (corresponding to the I- and Q- branches) to control the RIS. Fig.~\ref{Fig3}(f) depicts the transmitter structure of RIS-RQSM, where the RIS is divided into two halves that are controlled by the I- and Q- branches, respectively. In the absence of the $M$-PAM modulator, each data stream is divided into two parts that consist of $\log_2 N_R$ and 1 bits, respectively. As for the I- branch, the first part is used to determine the index of the targeted receive antenna corresponding to the I- branch; the second part with one bit is used to further determine whether the phase shift is $-\alpha_I$ or $-\alpha_I+ \pi$, where $\alpha_I$ denotes the phase of the channel associated with the targeted receive antenna corresponding to the I- branch. Similarly, as for the Q- branch, the first part is used to determine the index of the targeted receive antenna corresponding to the Q- branch; the second part with one bit is used to further determine whether the phase shift is $-\alpha_Q +\pi/2$ or $-\alpha_Q - \pi/2$, where $\alpha_Q$ denotes the phase of the channel associated with the targeted receive antenna corresponding to the Q- branch. Hence, the SE is $2(1+\log_2 N_R)$ bpcu. Fig.~\ref{Fig4}(c) shows an illustrative example of channel phase shifts with $N=2$ and $N_R=2$. In the presence of the $M$-PAM modulator, the Q- branch works as in the absence of the $M$-PAM modulator, while for the I- branch, since the PAM symbol itself introduces a phase shift of $\pi$, the RIS's phase shift is fixed to $-\alpha_I$ (i.e., $-\alpha_I + \pi$ is not applicable anymore) in order to avoid phase ambiguity, which results in a loss of 1-bit spatial information. In this case, the SE is $2\log_2 N_R +\log_2 M+1$ bpcu. Notably, the detection of the I- and Q- branches can be performed independently for RIS-RQSM.

\section*{Comparison and Performance Evaluation}
In this section, we make a comparison between antenna-based and metasurface-based modulation, and then evaluate their BER performance.
\begin{table}[!t]
	\caption{Comparison between antenna-based and metasurface-based modulation.}
    \label{Table 1}
	\centering
\begin{tabular}{|c|c|c|c|c|c|}
\hline
Class & Scheme     & \begin{tabular}[c]{@{}c@{}}Number of RF chains/\\ transmit antennas\end{tabular} & SE & TX/RX complexity & CSIT/CSIR \\ \hline \hline
\multirow{6}{*}{\begin{tabular}[c]{@{}c@{}}Antenna-based\\ modulation\end{tabular}} & MIMO       & $N_T/N_T$                                                                               & High   & High/high      & No/yes          \\ \cline{2-6}
                                                                                    & SM         & $1/N_T$                                                                               & Low   & Low/low      & No/yes          \\ \cline{2-6}
                                                                                    & Single-RF GSM        & $1/N_T$                                                                              & Medium   & Low/medium      & No/yes          \\ \cline{2-6}
                                                                                    & QSM        & $1/N_T$                                                                               & Medium   & Low/low      & No/yes          \\ \cline{2-6}
                                                                                    & RSM        & $N_T/N_T$                                                                               & Low   & High/low      & Yes/yes          \\ \cline{2-6}
                                                                                    & RQSM       & $N_T/N_T$                                                                                & Medium   & High/low      & Yes/yes          \\ \hline
\multirow{6}{*}{\begin{tabular}[c]{@{}c@{}}Metasurface-based\\ modulation\end{tabular}}      & RIS-MIMO   & $1/1$                                                                               & High    & Low/high      & No/yes          \\ \cline{2-6}
                                                                                    & MBM        & $1/1$                                                                               & Medium   & Low/medium      & No/yes          \\ \cline{2-6}
                                                                                    & RIS-SM/GSM & $1/1$                                                                               & Medium   & Low/medium      & Yes/yes          \\ \cline{2-6}
                                                                                    & RIS-QSM    & $1/1$                                                                               & Medium   & Low/medium      & Yes/yes           \\ \cline{2-6}
                                                                                    & RIS-RSM    & $1/1$                                                                                & Low   & Low/low      & Yes/yes          \\ \cline{2-6}
                                                                                    & RIS-RQSM   & $1/1$                                                                               & Medium   & Low/low      & Yes/yes          \\ \hline
\end{tabular}
\end{table}

Table~\ref{Table 1} summarizes the considered antenna-based and metasurface-based modulation schemes in terms of number of RF chains/transmit antennas, SE, TX/RX complexity, and CSI requirements. From Table~\ref{Table 1}, we evince that metasurface-based modulation is capable of transmitting information by using a single antenna and a single RF chain.

\begin{figure}[t]
	\centering
	\includegraphics[width=4.5in]{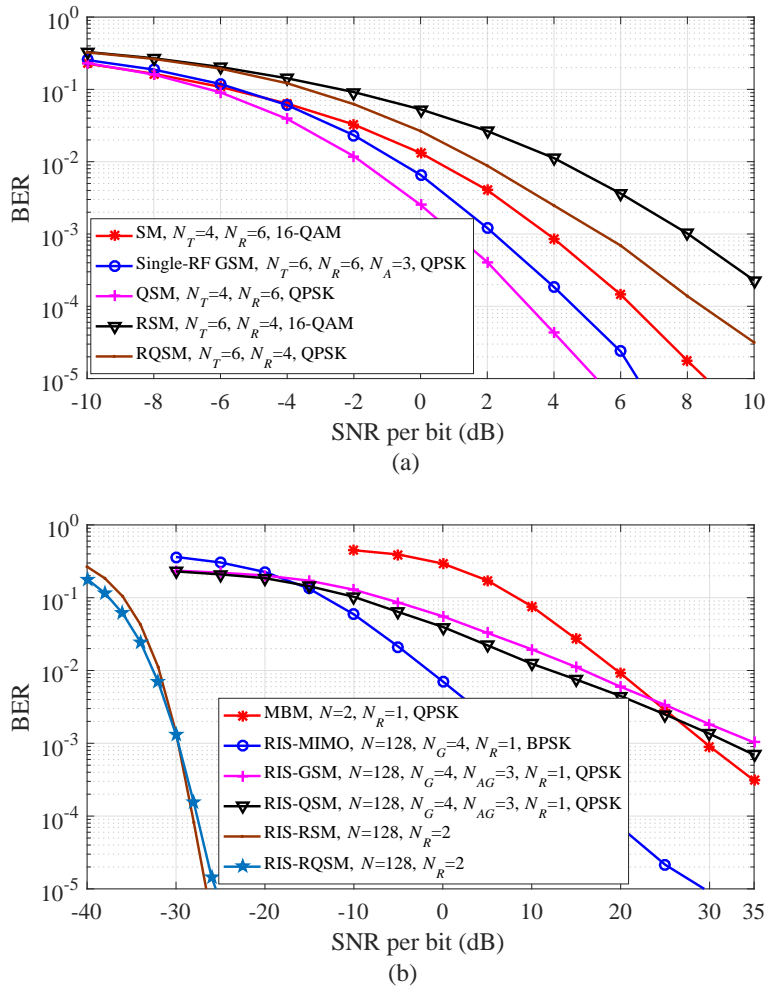}
	\caption{BER performance of (a) SM, single-RF GSM, QSM, RSM, and RQSM, and (b) MBM, RIS-MIMO, RIS-GSM, RIS-QSM, RIS-RSM, and RIS-RQSM with perfect CSIR over flat Rayleigh fading channels.}
	\label{Fig5}
\end{figure}

Fig.~\ref{Fig5} shows the BER of the considered antenna-based and metasurface-based modulation schemes. Specifically, in Fig.~\ref{Fig5}(a), we present the BER performance of SM, single-RF GSM, QSM, RSM, and RQSM with perfect CSI at the receiver (CSIR) and maximum-likelihood (ML) detectors over flat Rayleigh fading channels. The parameters for each scheme given in the legend are carefully chosen to achieve the same SE of 6 bpcu. RSM and RQSM adopt zero forcing-based pre-coding with perfect CSIT. Compared with SM and RSM, QSM and RQSM double the amount of spatial information, which in turn results in a smaller constellation size. At the BER value of $10^{-3}$, about 3 dB SNR gain can be achieved by QSM and RQSM over SM and RSM, respectively. In addition, at the expense of two more transmit antennas, single-RF GSM achieves about 1.5 dB SNR gain over SM.

In Fig.~\ref{Fig5}(b), we evaluate the BER of MBM, RIS-MIMO, RIS-GSM, RIS-QSM, RIS-RSM, and RIS-RQSM over flat Rayleigh fading channels. MBM, RIS-MIMO, RIS-GSM, and RIS-QSM employ one receive antenna and ML detection, while RIS-RSM and RIS-RQSM adopt two receive antennas and greedy detection. The number of RF mirrors is $N=2$ for MBM and that of RIS elements is $N=128$ for the other schemes. In RIS-MIMO, RIS-GSM, and RIS-QSM, the RIS elements are divided into $N_G$ groups, each containing $N/N_G$ elements with high channel correlation. For RIS-MIMO, each group transmits a common PSK/QAM symbol, while for RIS-GSM (RIS-QSM), $N_{AG}$ groups are activated (with the State-I) and the rest are inactive (with the State-II). In RIS-RSM and RIS-RQSM, we consider the case without $M$-ary symbols. As seen from Fig.~\ref{Fig5}(b), the considered metasurface-based modulation schemes vary remarkably in BER performance. RIS-RSM/RQSM significantly outperforms RIS-GSM/QSM at the cost of more receive antennas. This can be explained as follows. RIS-RSM/RQSM uses the indices of the receive antennas to convey spatial information. All RIS elements are exploited for beamforming so as to maximize the power received by the selected receive antenna, which is very favorable for the antenna index detection. However, RIS-GSM/QSM utilizes the indices of RIS elements to carry spatial information. Although beamforming is adopted at the RIS, it contributes less to the element index detection. Therefore, RIS-RSM/RQSM achieves a much higher reliability for the spatial information detection than RIS-GSM/QSM. Besides, the feasibilities of MBM and RIS-MIMO are verified in Fig.~\ref{Fig5}(b).

\section*{The Road Ahead}
In the previous section, the idea of metasurface-based single-RF modulation is illustrated via six paradigms. As a new emerging research field, there still exist numerous potential challenges and open research problems in metasurface-based single-RF modulation that deserve further investigations.

\subsection*{Hybrid Modulation}
The schemes discussed in the previous section are basic realizations of metasurface-based modulation. One extension is to investigate the scenario with the presence of a direct link, in which the reflection coefficients should be carefully designed by balancing the direct and reflecting links. Also, we can integrate some of the basic models to generate hybrid modulation schemes. For example, the idea of RIS-GSM that uses the indices of active elements for information transfer can be introduced into RIS-MIMO, RIS-QSM, RIS-RSM, and RIS-RQSM, which improves the SE by trading-off BER performance. With the aid of multiple antennas, SM-related techniques can be applied to the RF side in metasurface-based modulation. The transmit antennas at the RF side can also be replaced with reconfigurable antennas used in MBM.

\subsection*{Reflecting Element Grouping and Selection}
In metasurface-based modulation, several adjacent elements are usually grouped in order to facilitate the adjustment of the reflection coefficient. Since the grouping impacts the SE and BER performance, how to group the reflecting elements, such as the number of elements in each group, needs to be considered in the design of RIS-based systems. In RIS-QSM and RIS-RQSM, the selection of the elements for the I- and Q- branches also deserves further study. Moreover, in RIS-SM/GSM, since the value of ${{N}\choose{N_A}}$ may not be a power of two, an appropriate selection of the reflecting elements can be carried out to improve the BER performance.

\subsection*{Differential Modulation and Space-Time Coding}
Due to the large number of RIS elements, channel estimation is not an easy task for metasurface-based modulation based on coherent detection. Differential modulation techniques that can dispense with the need for CSI is an appealing research direction for RIS-based systems. Further, an RIS may be designed and configured to perform space-time coding on an incident signal to achieve diversity gains by reconfiguring the reflection coefficients.

\subsection*{Other Metasurface-Based Modulation Techniques}
Based on RISs, other modulation techniques can be invented. For instance, when number index modulation meets RISs, the number of active RIS elements can be varied according to the source data, rather than fixed as in RIS-SM/GSM. Besides, an RIS can manipulate an orthogonal frequency division multiplexing signal to enable subcarrier index modulation for transmitting its own information.

\section*{Conclusions}
In light of the challenges faced by SM, and motivated by the advantages of the RIS technology, we have proposed the idea of designing single-RF MIMO with RISs in this article. After revisiting the principle of SM, some representative SM-based implementations have been presented. Based on the principle of RISs, we have elaborated on some examples of metasurface-based single-RF MIMO. Finally, we have made a comparison between antenna-based and metasurface-based modulation, and evaluated their BER performance. Finally, we have discussed the challenges and opportunities on metasurface-based modulation. This article has shown that metasurface-based modulation provides a competitive alternative to antenna-based modulation for single-RF MIMO design.

%
%

\end{document}